\documentclass[aps,pre,epsf,superscriptaddress,amsmath,amssymb,amsfonts,twocolumn,showpacs]{revtex4-1}

\usepackage{graphicx}
\usepackage{epsfig}
\usepackage{dcolumn}
\usepackage{bm}
\usepackage{braket}
\usepackage{amsmath}
\usepackage{mathtools}
\usepackage{color}

\begin{document}
\title{Phase Separation Dynamics Induced by an Interaction Quench\\ of a Correlated Fermi-Fermi Mixture in a Double Well}

\author{J. Erdmann}
\affiliation{Zentrum f\"{u}r Optische Quantentechnologien,
Universit\"{a}t Hamburg, Luruper Chaussee 149, 22761 Hamburg,
Germany} 
\author{S. I. Mistakidis}
\affiliation{Zentrum f\"{u}r Optische Quantentechnologien,
Universit\"{a}t Hamburg, Luruper Chaussee 149, 22761 Hamburg,
Germany}
\author{P. Schmelcher}
\affiliation{Zentrum f\"{u}r Optische Quantentechnologien,
Universit\"{a}t Hamburg, Luruper Chaussee 149, 22761 Hamburg,
Germany} \affiliation{The Hamburg Centre for Ultrafast Imaging,
Universit\"{a}t Hamburg, Luruper Chaussee 149, 22761 Hamburg,
Germany}

\date{\today}

\begin{abstract}

We explore the interspecies interaction quench dynamics of ultracold spin-polarized few-body mass balanced Fermi-Fermi mixtures 
confined in a double-well with an emphasis on the beyond Hartree-Fock correlation effects. 
It is shown that the ground state of particle imbalanced mixtures exhibits a symmetry breaking of the single-particle 
density for strong interactions in the Hartree-Fock limit, which is altered within the many-body approach. 
Quenching the interspecies repulsion towards the strongly interacting regime the two species phase separate within the Hartree-Fock 
approximation while remaining miscible in the many-body treatment. 
Despite their miscible character on the one-body level the two species are found to be strongly correlated and exhibit a phase separation 
on the two-body level that suggests the anti-ferromagnetic like behavior of the few-body mixture. 
For particle balanced mixtures we show that an intrawell fragmentation (filamentation) of the density occurs both for the ground state 
as well as upon quenching from weak to strong interactions, a result that is exclusively caused by the presence 
of strong correlations. 
Inspecting the two-body correlations a phase separation of the two species is unveiled being a precursor towards an anti-ferromagnetic state.  
Finally, we simulate in-situ single-shot measurements and showcase how our findings can be retrieved by averaging over a sample of 
single-shot images.

\end{abstract}

\maketitle

\section{Introduction}

Ultracold Fermi gases offer an excellent testbed for simulating and exploring exotic quantum phases 
of matter \cite{giorgini2008theory,chevy2010ultra,bloch2008many}.  
Recent experimental advances constitute a valuable resource for disclosing the intricate complexity of 
condensed matter systems. 
Indeed several key quantities can be adjusted in the laboratory including the interparticle interaction strength via Feshbach 
resonances \cite{inouye1998observation,chin2010feshbach}, the particle number \cite{serwane2011deterministic,zurn2012fermionization,wenz2013few} 
and the external potential landscape \cite{bloch2008many,greiner2002quantum}. 
Besides single species also mixtures of fermions can nowadays be experimentally prepared 
with neutral fermionic atoms e.g. in different hyperfine states such as $\prescript{40}{}{K}$ \cite{wu2012ultracold, wille2008exploring}, 
$\prescript{6}{}{Li}$ \cite{wille2008exploring, moerdijk1995resonances} and $\prescript{87}{}{Sr}$ \cite{takamoto2006improved}.  

In this context, impressive features have been revealed evincing for instance superfluidity \cite{chen2005bcs, chin2006evidence}, 
quantum magnetism \cite{sowinski2018ground, hung2011exotic,yannouleas2016ultracold,koutentakis2018probing},  
insulating phases \cite{lisandrini2017topological, nataf2016chiral, zhou2016mott}, phase separation \cite{shin2006observation,partridge2006pairing,shin2008phase}, 
fermi polarons \cite {massignan2014polarons,scazza2017repulsive,schmidt2018universal,mistakidis2018repulsive} and Josephson 
junctions \cite{macri2013tunneling,valtolina2015josephson,spuntarelli2007josephson,erdmann2018correlated}. 
A major focus has been the phase diagram of Fermi-Fermi (FF) mixtures ranging from the strongly 
attractive to the strongly repulsive regime of interactions 
\cite{partridge2006pairing,iskin2007mixtures,cazalilla2009ultracold,jo2009itinerant, cazalilla2009ultracold, greif2013short, cherng2007superfluidity,sanner2012correlations}. 
For instance, referring to attractive particle imbalanced FF mixtures it has been shown that beyond a critical polarization the mixture forms a 
superfluid paired core being surrounded by a shell of unpaired fermions \cite{partridge2006pairing, iskin2007mixtures}. 
Turning to the repulsive regime of interactions magnetization effects emerge. 
For increasing repulsion, a first order phase transition \cite{cazalilla2009ultracold} between paramagnetism and 
itinerant ferromagnetism \cite{jo2009itinerant, cazalilla2009ultracold, greif2013short, cherng2007superfluidity,sanner2012correlations} has been revealed. 
It has been argued that this transition can be described by the mean-field model of Stoner \cite{stoner1933lxxx, snoke2009solid} 
for strongly short-range repulsively interacting fermions.

The majority of the above-mentioned studies has been focussed on the static properties of FF mixtures within a Hartree-Fock (HF) i.e. mean-field description 
in higher dimensions. 
Most importantly, the dynamical properties of FF mixtures are largely unexplored and especially the role of many-body (MB) effects is much less understood. 
An intriguing prospect here is whether magnetization or phase separation effects emerge during the nonequilibrium dynamics of FF mixtures. 
A widely used technique to induce the nonequilibrium dynamics is the so-called quantum quench \cite{polkovnikov2011colloquium,langen2015ultracold}, 
where the quantum evolution is generated following a sudden change of an intrinsic system's parameter such as the interaction strength 
\cite{kollath2007c,mistakidis2014interaction,mistakidis2015negative,mistakidis2017correlation,mistakidis2017mode}.  
For instance, it has been recently shown that the interaction quench dynamics of a Bose-Bose mixture crossing the miscibility-immiscibility 
threshold leads to the dynamical phase separation of the two clouds which exhibit domain-wall structures \cite{mistakidis2017correlation}.  
Turning to FF mixtures a natural question that arises is whether such a phase separation can be observed and what is its dependence on 
the particle number of each species \cite{ozawa2010population, combescot2001bcs}. 
Another interesting aspect here is whether any instabilities occuring in the HF approximation \cite{stoner1933lxxx} are altered due 
to the presence of correlations as well as the crucial role of the latter \cite{mazurenko2017cold,partridge2006pairing,sanner2012correlations,pekker2011competition} in the course of the evolution. 
Motivated also by the experimental capability to prepare few-fermion mixtures in one-dimension \cite{serwane2011deterministic,zurn2012fermionization,wenz2013few,murmann2015antiferromagnetic}, 
we study here the interaction quench dynamics of a spin-polarized FF mixture confined in a double-well. 
To simulate the correlated quantum dynamics of the FF mixture we employ the Multi-Layer Multi-Configurational Time-Dependent Hartree 
Method for Atomic Mixtures (ML-MCTDHX) \cite{ML-MCTDHX}, which is a variational method capturing all the important particle correlations. 

We find that the ground state of particle imbalanced species exhibits a symmetry breaking, for strong interactions, on the single-particle density level  
within the HF approximation. 
This behavior is a manifestation of the Stoner instability \cite{stoner1933lxxx, snoke2009solid} and renders the mixture immiscible. 
The presence of higher-order quantum correlations alters this instability and an intrawell fragmentation of the 
one-body density arises, i.e. the density profile breaks into several density branches (filaments) 
\footnote{Due to the presence of strong interspecies interactions the Gaussian-like density profile within each well deforms and either exhibits several local maxima or breaks 
into distinct density branches \cite{mistakidis2017correlation}. Throughout this work we refer to both these local maxima or density branches as filaments.}, while the two species remain miscible. 
Performing an interspecies interaction quench from weak-to-strong coupling we find that within the HF approximation the $\sigma$-species (with $\sigma=A,B$ denoting each species) 
single-particle density filamentizes and subsequently the two species phase separate. 
In sharp contrast, in the presence of quantum correlations the filamentation of the one-body density becomes suppressed 
and the fermionic components show a miscible behavior on the one-body level. 
Remarkably enough, Mott-like one-body correlations \cite{sherson2010single, larson2008mott,katsimiga2017dark,mistakidis2017correlation} 
between the filaments formed are revealed, indicating their tendency for localization. 
Most importantly, both the intra- and interspecies two-body correlation functions show that two fermions of the same or different species 
cannot populate the same filament but only distinct ones. 
The latter, which is arguably one of our main results, unveils that a phase separation process occurs only on the two-body 
level suggesting the formation of few-body anti-ferromagnetic like order \cite{murmann2015antiferromagnetic,koutentakis2018probing}.  

Turning to particle balanced FF mixtures we find that the single-particle density of the ground state exhibits a miscible behavior at weak and strong 
interactions in both the HF and MB approaches. 
Moreover, an intrawell fragmentation occurs only within the MB approach. 
Quenching the interspecies interaction from weak-to-strong coupling we observe that in the HF approximation 
the FF mixture remains miscible throughout the evolution, while performing an overall breathing motion. 
Within the MB approach the two species besides undergoing a breathing mode while remaining miscible, 
further exhibit an intrawell fragmentation (filamentation) of their single-particle density. 
Also in this case Mott-like one-body correlations appear between the distinct filaments formed. 
Moreover, two fermions of the same or different species exhibit an anti-correlated behavior in a single filament, 
whilst they are strongly correlated when residing in distinct filaments indicating the tendency towards an anti-ferromagnetic state. 
Finally, we simulate single-shot absorption measurements and showcase that by averaging a sample of in-situ images 
we can adequately reproduce the MB fermionic quench dynamics on the single-particle density level.  

This work is structured as follows. 
Section \ref{sec:theory} presents our setup and the basic observables of interest. 
The nonequilibrium dynamics induced by an interspecies interaction quench for particle imbalanced and balanced species within a double-well is analyzed 
in Secs. \ref{sec:particle_imbalanced} and \ref{sec:particle_balanced} respectively. 
We summarize our findings and provide an outlook in Section \ref{sec:conclusion}. 
In Appendix \ref{single_shots_details} we provide a brief discussion regarding our 
numerical implementation of the single-shot procedure. 
Finally, in Appendix \ref{sec:numerics} we present further details of our numerical simulations 
and demonstrate the convergence of the results discussed in the main text.

\section{Theoretical Framework}\label{sec:theory}

\subsection{Setup}\label{sec:setup} 

We consider a FF mixture consisting of $N_A$ and $N_B$ spin polarized fermions with equal masses $M_A=M_B\equiv M$ for the $A$ and $B$ species respectively. 
Such a mass balanced fermionic mixture can be experimentally realized by two different hyperfine states e.g. of $\prescript{40}{}{K}$ 
or $\prescript{6}{}{Li}$ \cite{wang2000ground, dieckmann2002decay}. 
These internal states could refer, for instance, to the $\ket{F=9/2,m_F-9/2}$ and $\ket{F=9/2,m_F=-7/2}$ of $\prescript{40}{}{K}$ \cite{zwierlein2006fermionic}. 
The mixture is confined in an one-dimensional double-well external potential \cite{bloch2005ultracold} which is composed by a harmonic 
oscillator with frequency $\omega$ and a centered Gaussian with height $V_0$ and width $w$. 
The resulting MB Hamiltonian reads 
\begin{align}
\mathcal{H}&=\sum\limits_{\sigma=A,B} \sum\limits_{i=1}^{N_\sigma}\left[  -\frac{\hbar^2}{2M}\left( \frac{d}{d x_i^\sigma}\right)^2
+\frac{1}{2}M\omega_\sigma^2(x_i^\sigma)^2 \right. \nonumber\\
&+ \left.\frac{V_0}{w \sqrt{2\pi}} e^{-\frac{(x_i^\sigma)^2 }{2w^2}}\right]\nonumber \\  &+ \sum\limits_{i=1}^{N_A} \sum \limits_{j=1}^{N_B}g_{AB}\delta(x_i^{A}-x_j^{B}).\label{Hamilt} 
\end{align} 
We operate in the ultracold regime, hence $s$-wave scattering is the dominant interaction process. 
Consequently the interspecies interactions can be adequately modeled by contact interactions, which scale with the 
effective one-dimensional coupling strength $g_{AB}$ for the different fermionic species. 
Since $s$-wave scattering is forbidden for spin-polarized fermions \cite{pethick2002bose, lewenstein2012ultracold}, due to the antisymmetry of the 
fermionic wavefunction, fermions of the same species are considered to be non-interacting. 
Therefore, only interspecies interactions are relevant in the MB Hamiltonian. 
The effective interspecies one-dimensional coupling strength \cite{olshanii1998atomic} is given by  
${g_{AB}} =\frac{{2{\hbar ^2}{a^s_{AB}}}}{{\mu a_ \bot ^2}}{\left( {1 - {\left|{\zeta (1/2)} \right|{a^s_{AB}}}/{{\sqrt 2 {a_ \bot }}}} \right)^{ -
1}}$, where $\zeta$ refers to the Riemann zeta function and $\mu=\frac{M}{2}$ is the corresponding reduced mass. 
${a_\bot } = \sqrt{\hbar /{\mu{\omega _ \bot }}}$ is the transversal length scale with transversal confinement frequency ${{\omega _ \bot }}$ 
and ${a^s_{AB}}$ is the three-dimensional $s$-wave scattering length between the two distinct species. 
We note that $g_{AB}$ can be experimentally adjusted either by means of ${a^s_{AB}}$ with the aid of Feshbach resonances \cite{kohler2006production, chin2010feshbach} 
or by manipulating ${{\omega _ \bot }}$ via confinement-induced resonances \cite{olshanii1998atomic, kim2006suppression}. 

In the following our Hamiltonian is rescaled in units of $\hbar  \omega_{\perp}$. 
Thus, the corresponding length, time, and interaction strength scales are expressed in terms of
$\sqrt{\frac{\hbar}{M \omega_{\perp}}}$, $\omega_{\perp}^{-1}$ and 
$\sqrt{\frac{\hbar^3 \omega_{\perp}}{M}}$ respectively. 
Moreover, the amplitude of the Gaussian barrier $V_0$, its width $w$ and the frequency of the harmonic oscillator $\omega$ 
are given in units of $\sqrt{ \frac{ \hbar^3 \omega_{\perp} }{ M }}$, $\sqrt{\frac{\hbar}{M \omega_{\perp}}}$, 
and $\omega_{\perp}$. 
To limit the spatial extension of our system we impose hard-wall boundary conditions at $x_\pm=\pm40$. 

Throughout this work, our system is initially prepared in the MB ground state of the Hamiltonian (\ref{Hamilt}) within the weak 
interspecies interaction regime, namely $g_{AB}=0.1$. 
The corresponding double-well potential is characterized by the harmonic oscillator frequency $\omega=0.1$, barrier height $V_0=2$ and width $w=1$. 
Thus, in a non-interacting single-particle picture four doublets are included below the maximum of the barrier. 
To induce the nonequilibrium dynamics of the FF mixture in the double-well we quench at $t=0$ the interspecies interaction strength towards the strongly 
correlated regime, e.g. $g_{AB}=4.0$, and let the system evolve in time. 
Quenching the interspecies repulsion towards the strongly interacting regime favors the occurrence of a breathing mode \cite{mistakidis2017correlation} and 
the appearance of strong intra- and interspecies correlations [see Secs. \ref{sec:particle_imbalanced} and \ref{sec:particle_balanced}] due to 
the quench imported interaction energy into the system \cite{fang2014quench}. 
Below, we first analyze the dynamics of a particle imbalanced mixture with $N_A=3$ ($N_A=5$) and $N_B=1$ fermions respectively, and subsequently examine 
the corresponding particle balanced case with $N_A=N_B=2$ and $N_A=N_B=5$.

\subsection{Many-Body Approach}\label{sec:wfn}

To solve the underlying MB Schr{\"o}dinger equation that governs the quech-induced dynamics of the FF mixture we 
utilize ML-MCTDHX \cite{ML-MCTDHX}. 
It is based on an expansion of the MB wavefunction with respect to a time-dependent and variationally optimized MB basis. 
Such a treatment enables us to take into account both the inter- and intraspecies correlations inherent in the system. 
In order to include the inter- and intraspecies correlations, we first introduce $M$ distinct species functions, 
$\Psi^{\sigma}_k (\vec x^{\sigma};t)$. 
Here, $\vec x^{\sigma}=\left( x^{\sigma}_1, \dots, x^{\sigma}_{N_{\sigma}} \right)$ refer to the spatial $\sigma=A,B$ species coordinates 
of each component consisting of $N_{\sigma}$ fermions. 
Then the MB wavefunction, $\Psi_{MB}$, is expressed as a truncated Schmidt 
decomposition \cite{horodecki2009quantum} of rank $D$ 
\begin{equation}
\Psi_{MB}(\vec x^A,\vec x^B;t) = \sum_{k=1}^D \sqrt{ \lambda_k(t) }~ \Psi^A_k (\vec x^A;t) \Psi^B_k (\vec x^B;t).    
\label{Eq:WF}
\end{equation} 
In this expression $D\le \min(\dim(\mathcal{H}^A),\dim(\mathcal{H}^B))$ and $\mathcal{H}^{\sigma}$ is the Hilbert space of the $\sigma$-species (see also the discussion below). 
The Schmidt coefficients $\lambda_k(t)$ in decreasing order are denoted as the natural species populations of the $k$-th 
species function $\Psi^{\sigma}_k$ of the $\sigma$-species. 
They serve as a measure of the system's entanglement or interspecies correlations.   
Specifically, the system is called entangled or interspecies correlated \cite{roncaglia2014bipartite} when at least two 
distinct $\lambda_k(t)$ are nonzero, since in this latter case the total MB state [Eq. (\ref{Eq:WF})] cannot be expressed 
as a direct product of two states. 

To explicitly incorporate the interparticle correlations each of 
the species functions $\Psi^{\sigma}_k (\vec x^{\sigma};t)$ is expanded using the determinants of $m^{\sigma}$ distinct 
time-dependent fermionic single-particle functions (SPFs), $\varphi_1,\dots,\varphi_{m_{\sigma}}$. 
In particular 
\begin{equation}
\begin{split}
&\Psi_k^{\sigma}(\vec x^{\sigma};t) = \sum_{\substack{l_1,\dots,l_{m_{\sigma}} \\ \sum l_i=N}} C_{k,(l_1,
	\dots,l_{m_{\sigma}})}(t)\\& \sum_{i=1}^{N_{\sigma}!} {\rm sign}(\mathcal{P}_i) \mathcal{P}_i
 \bigg[ \prod_{\substack{j\in\{1,\dots,m^{\sigma}\} \\ {\rm with}~ l_j=1}} \varphi_j(x_{K(j)};t)\bigg].  
\label{Eq:SPFs}
\end{split}
\end{equation} 
Here, $\mathcal{P}$ refers to the permutation operator which exchanges the particle positions $x_{\mu}$, $\mu=1,\dots,N_{\sigma}$ within 
the SPFs. 
Also $K(j)\equiv \sum_{\nu=1}^{j}l_{\nu}$ with $l_{\nu}$ denoting the occupation of the $\nu$th SPF and $j\in\{1,2,\dots,m^{\sigma}\}$. 
The symbol $\rm{sign}(\mathcal{P}_i)$ denotes the sign of the corresponding permutation and $C_{k,(l_1,\dots,l_{m_{\sigma}})}(t)$ 
are the time-dependent expansion coefficients of a certain determinant. 
The eigenfunctions of the one-body reduced density matrix of the $\sigma$-species 
$\rho_\sigma^{(1)}(x,x^\prime;t)=\langle\Psi_{MB}(t)|\hat{\Psi}^{\sigma,\dagger}(x)\hat{\Psi}^\sigma(x^\prime)|\Psi_{MB}(t)\rangle$ 
are termed natural orbitals $\phi^{\sigma}_i(x;t)$, 
where $\hat{\Psi}^{\sigma}(x)$ refers to the fermionic field operator of the $\sigma$-species. 
The eigenvalues of $\phi^{\sigma}_i(x;t)$ are the so-called natural populations $n^{\sigma}_i(t)$. 
If more than $N_\sigma$ natural populations, $n_i(t)$, possess a non-negligible occupation 
($0<n_i(t)<1$ with $N_{\sigma}<i<m^{\sigma}$), the fermionic $\sigma$-species is termed intraspecies correlated, otherwise the 
MB state reduces to the HF ansatz \cite{pethick2002bose,giorgini2008theory,pitaevskii2016bose}.  
Indeed ML-MCTDHX enables us to operate within different approximation 
orders \cite{ML-MCTDHX}, and we e.g. retrieve the HF ansatz \cite{pethick2002bose} in the limit 
of $D=1$ and $m^{\sigma}=N_{\sigma}$ 
\begin{equation}
\begin{split}
&\Psi_{HF}(\vec x^{A},\vec x^{B};t) =\\ 
	&\prod_{\sigma=A,B}\sum_{i=1}^{N_{\sigma}!} {\rm sign}(\mathcal{P}_i) \mathcal{P}_i
\left[\varphi_1(x_1^{\sigma};t) \cdots \varphi_{N_{\sigma}}(x_{N_{\sigma}}^{\sigma};t) \right].  
\label{Eq:HF}
\end{split}
\end{equation} 
Furthermore, employing the Dirac-Frenkel variational principle \cite{frenkel1932wave,dirac1930note} for the MB ansatz 
[see Eqs.~(\ref{Eq:WF}), (\ref{Eq:SPFs})] we obtain the ML-MCTDHX equations of motion \cite{ML-MCTDHX} for the fermionic mixture. 
These equations correspond to $D^2$ linear differential equations of motion for the coefficients $\lambda_i(t)$ coupled to a set of 
$D$[${m_A}\choose{N_A}$+$ {m_B}\choose{N_B}$] non-linear integro-differential equations for the species functions and $m^A+m^B$  
integro-differential equations for the SPFs.

\subsection{Correlation Functions}\label{def_cor_functions} 

To unveil the degree of intraspecies correlations at the one-body level during the quench dynamics we employ the 
normalized spatial first order correlation function \cite{naraschewski1999spatial, sakmann2008reduced,mistakidis2017correlation}
\begin{align}
g^{(1)}_\sigma(x,x^\prime;t)=\frac{\rho_\sigma^{(1)}(x,x^\prime;t)}{\sqrt{\rho_\sigma^{(1)}(x;t)\rho_\sigma^{(1)}(x^\prime;t)}}.\label{one_body_cor}
\end{align} 
Here, $\rho_\sigma^{(1)}(x,x^\prime;t)=\langle\Psi(t)|\hat{\Psi}^{\sigma,\dagger}(x)\hat{\Psi}^\sigma(x^\prime)|\Psi(t)\rangle$ refers to the 
one-body reduced density matrix of the $\sigma$ species and $\rho_\sigma^{(1)}(x;t)\equiv\rho_\sigma^{(1)}(x,x^\prime=x;t)$ is the one-body density. 
$\hat{\Psi}^{\sigma,\dagger}(x)$ [$\hat{\Psi}^{\sigma}(x)$] is the fermionic field operator that creates [annihilates] a $\sigma$ species 
fermion at position $x$. 
$|g^{(1)}_{\sigma}(x,x';t)|$ is bounded within the interval $[0,1]$ and measures the proximity of the MB state to a product 
state for a fixed set of coordinates $x$, $x'$. 
Two different spatial regions $R$, $R'$, with $R \cap R' = \varnothing$, possessing $|g^{(1)}_{\sigma}(x,x';t)|= 0$ with $x\in R$ and $x'\in R'$ are 
referred to as perfectly incoherent, whilst for $|g^{(1)}_{\sigma}(x,x';t)|= 1$, $x\in R$, $x'\in R'$ the regions are said to be fully coherent. 
When at least two distinct spatial regions are partially incoherent, i.e. $|g^{(1)}_{\sigma}(x,x';t)|<1$ this signifies the 
emergence of one-body intraspecies correlations, while their absence is designated by $|g^{(1)}_{\sigma}(x,x';t)|=1$ 
for every $x$, $x'$. 
Most importantly, the situation where a certain spatial region $R$ is fully coherent, i.e. $|g^{(1),\sigma}(x,x';t)|^2 \approx 1$ $x,x'\in R$, 
and perfect incoherence occurs between different spatial regions $R$, $R'$, i.e. $|g^{(1),\sigma}(x,x';t)|^2\approx 0$, $x\in R$, $x'\in R'$ 
with $R \cap R' = \varnothing$, indicates the appearance of Mott-like correlations \cite{sherson2010single, larson2008mott,katsimiga2017dark,mistakidis2017correlation}.  

To estimate the degree of second order intra- and interspecies correlations in the course of the dynamics, we inspect the normalized 
two-body correlation function \cite{sakmann2008reduced,mistakidis2017correlation}
\begin{align}
g^{(2)}_{\sigma\sigma\prime}(x,x^\prime;t)=\frac{\rho_{\sigma\sigma\prime}^{(2)}(x,x^\prime;t)}{\rho_\sigma^{(1)}(x;t)\rho_{\sigma\prime}^{(1)}(x^\prime;t)}. \label{two_body_cor}
\end{align}
In Eq. (\ref{two_body_cor}), $\rho^{(2)}(x,x';t)=\langle\Psi(t)|\hat{\Psi}^{\sigma,\dagger}(x')\hat{\Psi}^{\dagger,\sigma^\prime}(x)\hat{\Psi}^{\sigma^\prime}
(x)\\\hat{\Psi}^\sigma(x')|\Psi(t)\rangle$ denotes the diagonal two-body reduced density matrix which provides the probability of measuring two particles of 
species $\sigma$ and $\sigma^\prime$ located at $x$ and $x^\prime$ respectively at time $t$. 
Referring to the same (different) species, i.e. $\sigma=\sigma'$ ($\sigma \not=\sigma'$), $|g^{(2)}_{\sigma \sigma'}(x,x^\prime;t)|$ 
accounts for the intraspecies (interspecies) two-body correlations. 
We remark that if $g^{(2)}_{\sigma\sigma\prime}(x,x^\prime;t)=1$ holds, the state is termed fully second order coherent, 
while in case that $g^{(2)}_{\sigma\sigma\prime}(x,x^\prime;t)>1$ [$g^{(2)}_{\sigma\sigma\prime}(x,x^\prime;t)<1$] it is 
termed correlated [anti-correlated]. 
$g^{(2)}_{\sigma\sigma\prime}(x,x^\prime;t)$ is experimentally accessible by in-situ density-density fluctuation 
measurements \cite{tavares2016chaotic, tsatsos2017granulation, endres2013single}.

\section{Interaction Quench Dynamics of a Particle Imbalanced Mixture}\label{sec:particle_imbalanced}

\begin{figure}
	\includegraphics[trim=0 0 0 0,clip,width=0.48\textwidth]{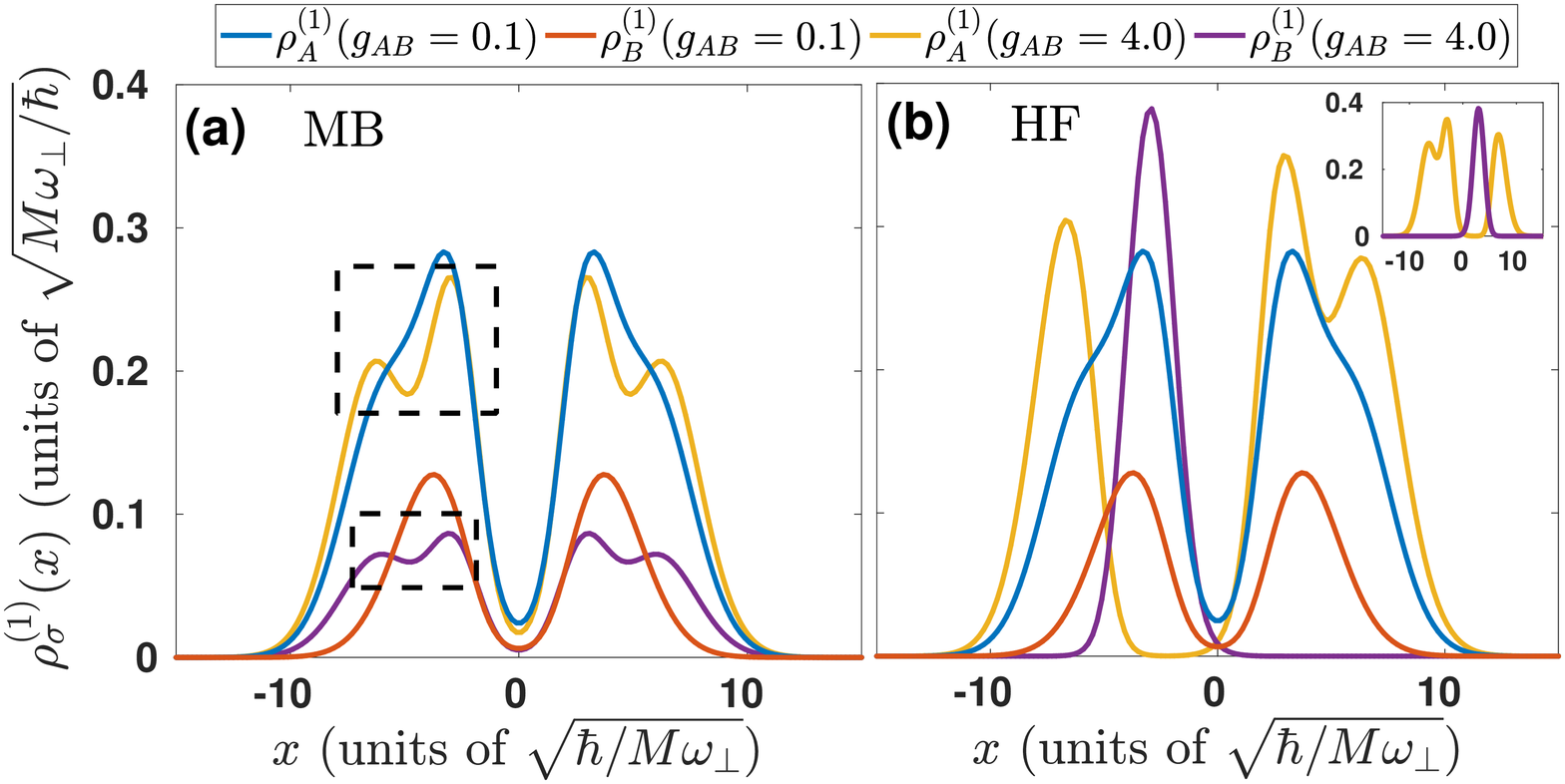}
	\caption{One-body density $\rho^{(1)}_\sigma(x)$ of the ground state of the $\sigma$-species of a FF mixture for different 
	interspecies repulsions $g_{AB}$ (see legend) within (a) the MB approach and (b) the HF approximation. 
	The mixture consists of $N_A=3$ and $N_B=1$ fermions and it is trapped in a double-well potential. 
	The rectangles in (a) indicate the intrawell fragmentation (filamentation) of $\rho^{(1)}_\sigma(x)$ 
	occuring for strong interspecies interactions. 
	The inset in (b) shows the corresponding energetically degenerate configuration of $\rho^{(1)}_\sigma(x)$ with the $\rho^{(1)}_\sigma(x)$ 
	of the main panel for $g_{AB}=4.0$ within the HF approximation.}
	\label{abb:gs3-1} 
\end{figure}

\subsection{Initial State}\label{in_par_imb}

We consider an interspecies repulsively interacting ($g_{AB}$) mass balanced ($M_A=M_B$) FF mixture with spin-polarized and particle imbalanced components 
consisting of $N_A=3$ and $N_B=1$ fermions. 
The mixture is confined within a double-well and it is initialized in its corresponding interspecies 
interacting ground state as described by the Hamiltonian of Eq. (\ref{Hamilt}), using either imaginary time propagation or improved relaxation \cite{ML-MCTDHX} within ML-MCTDHX. 
The double-well is characterized by frequency $\omega=0.1$, barrier height $V_0=2$ and width $w=1$. 

To inspect the ground state of the FF mixture we invoke the $\sigma$-species single-particle density 
$\rho_\sigma^{(1)}(x;t)$ [see also Eq. (\ref{one_body_cor})]. 
Within the weakly interacting regime, $g_{AB}=0.1$, we observe that each $\rho_\sigma^{(1)}(x)$ shows an equal population in the two wells 
of the double-well and it is distributed in a symmetric manner both in the HF approximation as well as on the 
MB level, see Figs. \ref{abb:gs3-1} (a) and (b). 
Also, $\rho_A^{(1)}(x;t)$ and $\rho_B^{(1)}(x;t)$ feature a miscible behavior in both approaches. 
Note that due to the particle imbalance the $A$-species which contains the higher particle number exhibits a broader single-particle density distribution 
within each well when compared to the $B$-species. 
Turning to the strong interaction regime, $g_{AB}=4.0$, an intrawell fragmentation of $\rho_\sigma^{(1)}(x)$ occurs on 
the MB level, see Fig. \ref{abb:gs3-1} (a), while $\rho_A^{(1)}(x;t)$ and $\rho_B^{(1)}(x;t)$ show again a miscible behavior. 
Intrawell fragmentation refers to the filamentation tendency of the one-body density, i.e. to the appearance of several local maxima 
occurring in $\rho_\sigma^{(1)}(x)$ within each well, see the dashed rectangle in Fig. \ref{abb:gs3-1} (a) where two sub maxima are present. 
Interestingly enough, within the HF approximation the mixture becomes immiscible with $\rho_A^{(1)}(x)$ and $\rho_B^{(1)}(x)$ being phase 
separated as it can be observed by their asymetric distribution with respect to $x=0$ illustrated in Fig. \ref{abb:gs3-1} (b). 
This latter behavior can be thought of as the few-body analog of the Stoner's instability \cite{jacquod2000supression, jacquod2001ground} which is a well-known 
phenomenon in solid state physics being responsible for magnetization effects emerging in itinerant systems. 
Indeed, within the HF approximation in the strongly interacting regime the energy of a miscible state is larger 
when compared to the energy of a phase separated (immiscible) one due to the strong impact of the interaction energy \cite{stoner1933lxxx,snoke2009solid}. 
Thus, the particle number assymetry favors a phase separated state with $\rho_B^{(1)}(x)$ being localized 
in one of the wells and $\rho_A^{(1)}(x)$ distributing around it, see Fig. \ref{abb:gs3-1} (b). 
Since the double-well is symmetric, the same occupation structure of $\rho_A^{(1)}(x)$ and $\rho_B^{(1)}(x)$ with interchanged wexlls possesses 
an equal energy, i.e. the two configurations are energetically degenerate, see the inset of Fig. \ref{abb:gs3-1} (b). 
Recall that this phenomenon occurs, in the one-dimensional spin-independent case considered here, only within the HF approximation and not at the MB level in accordance to the 
Lieb-Mattis theorem \cite{lieb1962lieb}. 
In the latter approach the Stoner 
instability ceases to exist due to the involvement of higher-order quantum superpositions \cite{koutentakis2018probing,volosniev2014strongly}.  

Next, we examine the quantum dynamics of the above-mentioned weakly interacting, $g_{AB}=0.1$, FF mixture by quenching the interspecies repulsion at $t=0$ 
towards the strongly correlated regime of interactions, $g_{AB}=4.0$.  

\begin{figure}
	\includegraphics[trim=0 0 0 0,clip,width=0.49\textwidth]{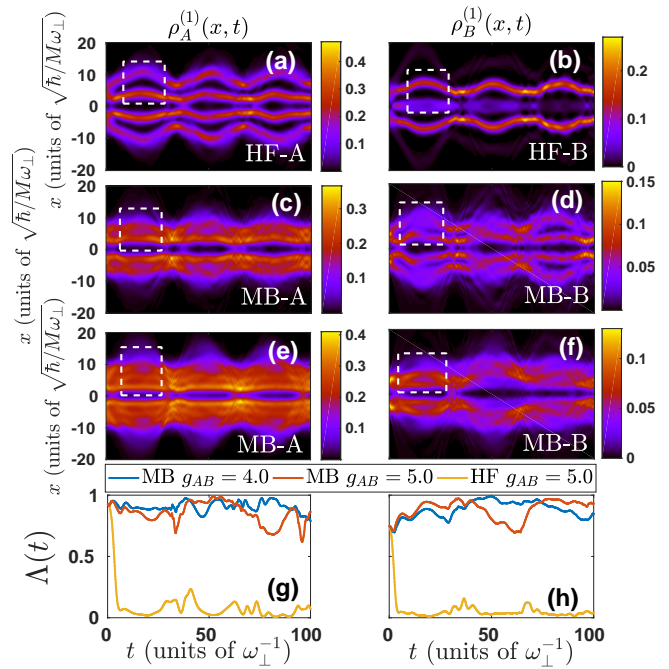}
	\caption{Evolution of the $\sigma$-species one-body density $\rho^{(1)}_\sigma(x;t)$ of a FF mixture within the (a), (b) HF approximation and (c)-(f) MB approach 
	following an interaction quench from $g_{AB}=0.1$ to $g_{AB}=4.0$. 
	The FF mixture consists of (a)-(d) $N_A=3$ and $N_B=1$ particles and (e), (f) $N_A=5$ and $N_B=1$ fermions. 
	The left and right columns correspond to the densities of the $A$ and the $B$ species respectively. 
	The rectangles indicate the number of filaments formed of the corresponding $\rho^{(1)}_\sigma(x;t)$ within 
	the left well. 
	(g), (h) Overlap integral $\Lambda(t)$ between the species of the FF mixture during the evolution within different approximations and varying postquench 
	interactions (see legend) for (g) $N_A=3$, $N_B=1$ and (h) $N_A=5$, $N_B=1$. }
	\label{abb:den3-1} 
\end{figure}

\subsection{Single-Particle Density Evolution}\label{one_body_par_imb}

To visualize the nonequilibrium dynamics of the particle imbalanced FF mixture on the one-body level we employ the single-particle density 
evolution $\rho_\sigma^{(1)}(x;t)$ for each of the species after the quench. 
Focusing on the HF approximation, see Figs. \ref{abb:den3-1} (a) and (b), we observe that an overall breathing mode \cite{abraham2012quantum,abraham2014quantum,abraham2014quantum} 
of both fermionic clouds takes place manifested as a contraction and expansion dynamics of $\rho_\sigma^{(1)}(x;t)$. 
The frequency of this breathing mode is $\omega_{br}=0.2=2\omega$ which is in accordance with the corresponding theoretical prediction \cite{abraham2014quantum}. 
Most importantly, a phase separation process between the two species occurs and each $\rho_\sigma^{(1)}(x;t)$ exhibits an intrawell fragmentation. 
This phase separation is a consequence of the Stoner instability that exists in this strongly interacting regime even in the ground state of 
the system [see also our discussion in Sec. \ref{in_par_imb}.] 
Regarding the intrawell fragmentation we observe that $\rho_A^{(1)}(x;t)$ forms two filaments in each well, 
while $\rho_B^{(1)}(x;t)$ exhibits one filament in each well and one (of lower amplitude) located at the position of the barrier 
of the double-well [see also the dashed rectangles in Figs. \ref{abb:den3-1} (a) and (b)].  
In sharp contrast to the above, utilizing the correlated approach the single-particle density evolution shows a completely 
different behavior, see Figs. \ref{abb:den3-1} (c) and (d). 
The two components remain miscible throughout the evolution in accordance to the ground state properties discussed in Sec. \ref{in_par_imb}. 
Moreover, an intrawell fragmentation emerges with the two filaments formed in $\rho_\sigma^{(1)}(x;t)$ within each well being more pronounced 
for the $B$-species, while the filamentary structure of the $A$-species is suppressed and hardly discernible [see also the rectangles in Figs. \ref{abb:den3-1} (c) and (d)]. 
Finally, both clouds undergo a breathing motion with approximately the same frequency as the one observed in the HF approximation.  

To infer about the effect of the majority species particle number on the nonequilibrium dynamics, we next consider a mass balanced FF mixture 
with $N_A=5$ and $N_B=1$. 
The corresponding single-particle density evolution following an interspecies interaction quench from $g_{AB}=0.1$ to $g_{AB}=4.0$ 
is shown in Figs. \ref{abb:den3-1} (e) and (f) within the MB approach. 
As it can be deduced, a larger particle number of the majority component leads to an increased number of filaments within each well for each of the species 
as compared to the case of a smaller particle number [compare Figs. \ref{abb:den3-1} (c), (d) and (e), (f)]. 
It is also worth mentioning that the filament formation of both species is washed out for higher particle numbers [see also the rectangles in Figs. \ref{abb:den3-1} (e) and (f)]. 
However, the particle number does not significantly alter the breathing frequency of each species and their miscible character. 

To expose the degree of spatial phase separation, namely the degree of miscibility or immiscibility of the mixture, 
occuring on the one-body level during the quench dynamics, we employ the overlap integral function 
$\Lambda(t)$ \cite{mistakidis2017correlation, bandyopadhyay2017dynamics, jain2011quantum} between the two species 
\begin{align}
\Lambda (t)=\frac{\left[\int d x \, \rho_A^{(1)}(x;t)\rho_B^{(1)}(x;t)\right]^2}{\left[\int d x\, (\rho_A^{(1)}(x;t))^2\right]\left[\int d x\, (\rho_B^{(1)}(x;t))^2\right]}.
\end{align} 
This quantity being normalized to unity takes values between $\Lambda=0$ and $\Lambda=1$ corresponding to zero and complete spatial overlap of the two species 
on the single-particle level. 
Figures \ref{abb:den3-1} (g) and (h) present $\Lambda(t)$ for the setups $N_A=3$, $N_B=1$ and $N_A=5$, $N_B=1$ respectively for different 
interaction quench amplitudes. 
Regarding the evolution in the HF approximation, $\Lambda(t)$ drops close to zero at short time scales ($t>6$) for both systems. 
After this initial drop the overlap remains almost constant exhibiting small amplitude oscillations which reflect the breathing motion of each cloud. 
Notice that the maxima of these small amplitude oscillations appear at time intervals of the contraction of the 
cloud, see e.g. Figs. \ref{abb:den3-1} (a) and (g) at $t\approx40$. 
In sharp contrast to the above behavior, $\Lambda(t)$ shows small fluctuations around 0.9 within the MB approach during the entire evolution. 
The aforementioned evolution of $\Lambda(t)$ reflects the miscible character of the dynamics on the single-particle level. 
The same overall phenomenology in terms of $\Lambda(t)$ holds equally, in both approaches, for other postquench interaction strengths, see e.g. 
Figs. \ref{abb:den3-1} (g) and (h) for $g_{AB}=4.0$. 
We further remark that for postquench interaction strengths $g_{AB}>2.0$ the overlap function features a similar dynamics, while 
for quenches to $g_{AB}<2.0$ the mixture remains miscible in both the HF and the MB approach (results not shown here for brevity). 

\begin{figure}
	\includegraphics[trim=0 0 50 0,clip,width=0.49\textwidth]{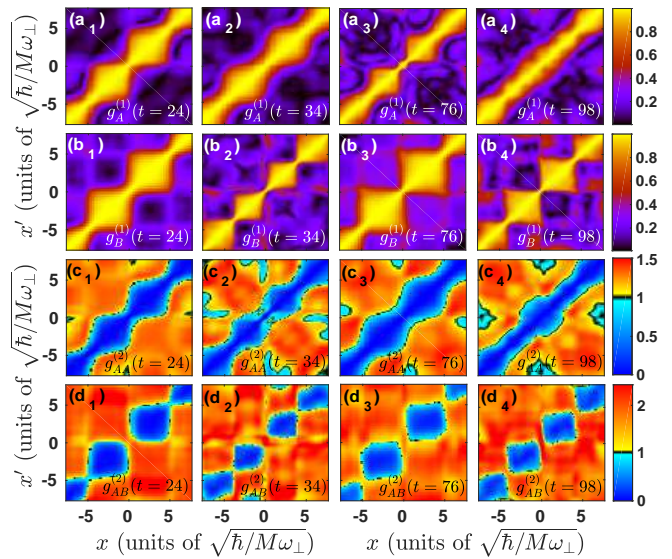}
	\caption{One-body correlation function $g_\sigma^{(1)}(x,x^\prime;t)$ shown for different time instants (see legends) during the interaction quench dynamics 
	of a FF mixture for ($a_1$)-($a_4$) the $A$-species and ($b_1$)-($b_4$) the $B$-species. 
	($c_1$)-($c_4$) Snapshots of the two-body intraspecies correlation function $g_{AA}^{(2)}(x,x^\prime;t)$ and ($d_1$)-($d_4$) the interspecies 
	two-body correlation function $g_{AB}^{(2)}(x,x^\prime;t)$.  
	In all cases, the FF mixture consists of $N_A=3$, $N_B=1$ particles and it is initialized in the weakly interacting ground state, $g_{AB}=0.1$, of the double-well.  
	To induce the dynamics we perform an interaction quench from $g_{AB}=0.1$ to $g_{AB}=4.0$. } 
	\label{abb:coh3-1} 
\end{figure}

\subsection{Correlation Dynamics}\label{cor_par_imb}

To unveil the underlying correlation mechanisms \cite{sanner2012correlations,pekker2011competition} that lead to the intrawell fragmentation during the MB quench dynamics, we investigate 
the one-body $g_{\sigma}^{(1)}(x,x^\prime,t)$ [Eq. (\ref{one_body_cor})] and the two-body $g_{\sigma\sigma'}^{(2)}(x,x',t)$ [Eq. (\ref{two_body_cor})] 
intra- and interspecies correlation functions during evolution, see Fig. \ref{abb:coh3-1}. 
As it is expected the intraspecies two-body correlation function for the $B$-species is zero, since this species contains only a single particle. 
Below we examine $g_{\sigma}^{(1)}(x,x^\prime,t)$ and $g_{\sigma\sigma'}^{(2)}(x,x',t)$ following an interaction quench of the FF mixture 
from $g_{AB}=0.1$ to $g_{AB}=4.0$. 

Figures \ref{abb:coh3-1} (a\textsubscript{1})-(a\textsubscript{4}) and (b\textsubscript{1})-(b\textsubscript{4}) show $g_{A}^{(1)}(x,x^\prime,t)$ and 
$g_{B}^{(1)}(x,x^\prime;t)$ respectively for selected time instants of the MB evolution. 
Overall, we observe that throughout the evolution the off-diagonal elements of $g_{\sigma}^{(1)}(x,x^\prime,t)$ are supressed. 
Indeed on the one-body level, each filament of both species is fully coherent with itself 
[see e.g. $g_A^{(1)}(x=-2.5,x^\prime=-2.5;t=24)\approx1$ in Fig. \ref{abb:coh3-1} (a\textsubscript{1})] and mainly incoherent with 
any of the other filaments [see e.g. $g_A^{(1)}(x=-2.5,x^\prime=2.5;t=24)\approx0$ in Fig. \ref{abb:coh3-1} (a\textsubscript{1})]. 
We note that this behavior of $g_{\sigma}^{(1)}(x,x^\prime,t)$ is more pronounced for the $A$-species, while in the $B$-species two distinct filaments 
appear to be partially incoherent [e.g. $g_B^{(1)}(x=-2.5,x^\prime=2.5;t=76)\approx0.3$ in Fig. \ref{abb:coh3-1} (b\textsubscript{3})], 
as shown in Figs. \ref{abb:coh3-1} (b\textsubscript {1})-(b\textsubscript{4}). 
This structure of $g_{\sigma}^{(1)}(x,x^\prime,t)$ indicates the occurrence of Mott-like correlations \cite{sherson2010single, larson2008mott,katsimiga2017dark} 
in the system [see also the corresponding discussion in Sec. \ref{def_cor_functions}] and suggests the tendency of the observed 
filaments to be localized structures. 
Moreover, we can infer that each $\sigma$-species fermion, and especially the $A$-species ones, is more likely to be localized in one filament 
and do not reside in two or more filaments. 

We next study the two-body intraspecies correlation function $g_{AA}^{(2)}(x,x';t)$, see Figs. \ref{abb:coh3-1} (c\textsubscript{1})-(c\textsubscript{4}). 
A strongly anti-correlated behavior within each filament, see the depleted diagonal behavior, occurs for every time instant 
[e.g. $g_{AA}^{(2)}(x=-2.5,x^\prime=-2.5;t=24)\approx0$ in Fig. \ref{abb:coh3-1} (c\textsubscript{1})], while   
two different filaments appear to be correlated [e.g. $g_{AA}^{(2)}(x=-2.5,x^\prime=2.5;t=24)\approx1.3$ in Fig. \ref{abb:coh3-1} (c\textsubscript{1})]. 
As a consequence, two particles of the $A$-species cannot reside in the same filament but they are more likely to be found in any pair of distinct filaments. 
The corresponding interspecies correlation function $g_{AB}^{(2)}(x,x')$, shown in Figs. \ref{abb:coh3-1} (d\textsubscript{1})-(d\textsubscript{4}), displays similar 
characteristics to $g_{AA}^{(2)}(x,x';t)$. 
Namely, a correlation hole exists [see e.g. $g_{AB}^{(2)}(x=-2.5,x^\prime=-2.5;t=24)\approx0$ in Fig. \ref{abb:coh3-1} (d\textsubscript{1})] 
which excludes the possibility of an $A$ and a $B$ particle to be in the same filament. 
However, the off-diagonal elements of $g_{AB}^{(2)}(x,x';t)$ exhibit a correlated behavior [see e.g. $g_{AB}^{(2)}(x=-2.5,x^\prime=2.5;t=24)\approx2.2$ 
in Fig. \ref{abb:coh3-1} (d\textsubscript{1})] providing the possibility for an $A$ and a $B$-species fermion to be located at different filaments. 
We remark, that similar correlation structures have also been observed for the $N_A=5$, $N_B=1$ case (not shown here for brevity). 

An important conclusion that can be extracted from the above analysis is that on the MB level phase separation between the species 
can be inferred only on the two-body level and not by simply observing the corresponding single-particle densities. 
Recall that the single-particle density evolution does not exhibit any phase separation within the MB 
approach, see Figs. \ref{abb:den3-1} (c), (d), which is in sharp contrast to the HF approximation where 
the fermionic components are evidently immiscible, see Figs. \ref{abb:den3-1} (a), (b). 
Combining also the results of $g_{\sigma}^{(1)}(x,x^\prime,t)$ and $g_{\sigma\sigma'}^{(1)}(x,x^\prime,t)$, it becomes 
apparent that all $N_A+N_B$ fermions reside in distinct filaments. 
Therefore regarding the spatially resolved distribution of the system, a superposition state consisting of all permutations of 
possible fermionic configurations concerning the four filaments formed, i.e $(B-A-A-A)$, $(A-B-A-A)$, $(A-A-B-A)$ and $(A-A-A-B)$, is permitted. 
This latter behavior suggests the tendency towards an anti-ferromagnetic like state of the few-body system \cite{koutentakis2018probing,murmann2015antiferromagnetic}. 

\begin{figure}
	\includegraphics[width=0.49\textwidth]{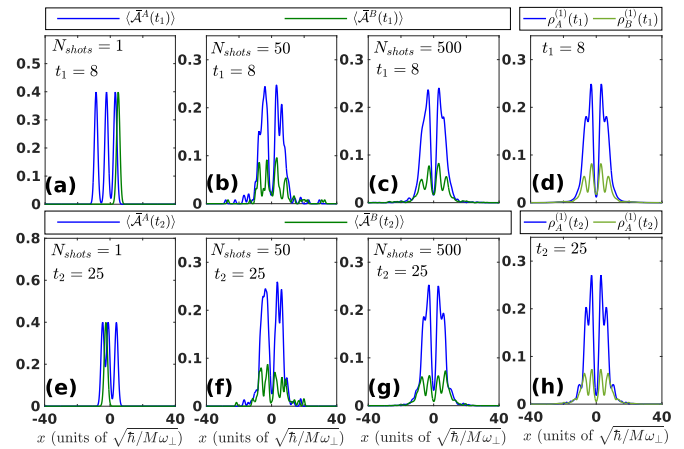}
	\caption{Single-shot images of each species, at distinct time instants of the interaction quench dynamics (see legends), 
	obtained by averaging over (a), (e) $N_{shots}=1$, (b), (f) $N_{shots}=50$, and (c), (g) $N_{shots}=500$. 
	(d), (h) The corresponding $\rho^{(1)}_{\sigma}(t)$ is evaluated within the MB approach. 
	In all cases the system consists of $N_A=3$ and $N_B=1$ fermions confined in a double-well potential and the dynamics is induced by quenching 
	the interspecies interaction from $g_{AB}=0.1$ to $g_{AB}=4.0$.}
	\label{abb:shots3-1} 
\end{figure}

\subsection{Single-Shot Images}\label{shots_par_imb}

In order to offer further possible experimental evidences of the correlated quantum quench dynamics of the FF mixture 
we simulate in-situ single-shot absorption measurements \cite{sakmann2016single,mistakidis2017correlation,lode2017fragmented}. 
These measurements essentially probe the spatial configuration of the atoms and they are based on the MB probability distribution which is 
available within ML-MCTDHX \cite{ML-MCTDHX}. 
To simulate the corresponding experimental procedure we perform a convolution of the spatial particle configuration with a point spread function 
being determined by the corresponding experimental resolution. 
For more details regarding the numerical implementation of this procedure in binary systems we refer the 
interested reader to Appendix \ref{single_shots_details}, while more elaborated discussions are provided in Refs. \cite{mistakidis2017correlation,erdmann2018correlated}.  
The point spread function used here possesses a Gaussian shape with width $w_{PSF}=1 \ll l\approx 3.2$, 
where $l=\sqrt{1/\omega}$ denotes the corresponding harmonic oscillator length. 
We note that in few-body experiments \cite{zurn2012fermionization,wenz2013few} fluorescence imaging is another promising technique 
to probe the state of the system since it eliminates unavoidable noise sources that might destroy the 
experimental signal \cite{serwane2011deterministic}. 
However, the simulation of this experimental technique lies beyond our current scope. 
Here, by simulating single-shot measurements we aim to show how in-situ imaging can be used to adequately monitor 
the nonequilibrium quantum dynamics of the aforementioned few-body particle imbalanced FF mixture. 

Utilizing the MB wavefunction of the system, obtained within ML-MCTDHX, we simulate in-situ single-shot images at each time instant $t$ of the MB evolution. 
Consecutively imaging first the $A$ and then the $B$ species at time $t\equiv t_{im}$, these images are designated by $\mathcal{A}^A(\tilde{x};t_{im})$ 
and $\mathcal{A}^B(\tilde{x'}|\mathcal{A}^A(\tilde{x});t_{im})$ for the $A$ and $B$ species respectively. 
In the following, we focus on the dynamics of a FF mixture with $N_A=3$ and $N_B=1$ within the double-well upon quenching 
the interspecies interaction strength from $g_{AB}=0.1$ to $g_{AB}=4.0$. 
Figures \ref{abb:shots3-1} (a), (e) show the first simulated in-situ single-shot images for each species, $\mathcal{A}^A(\tilde{x};t)$ 
and $\mathcal{A}^B(\tilde{x'}|\mathcal{A}^A(\tilde{x});t)$, at two distinct time instants during evolution, namely at $t_1=8$ and $t_2=25$. 
As it can be seen the images for both species, and especially the $\mathcal{A}^A(\tilde{x};t)$, exhibit a filamentized structure resembling this way 
the overall tendency observed in the one-body density evolution, see also Figs. \ref{abb:den3-1} (c), (d). 
Moreover, let us comment that the spatial position of these images is in accordance with our previous discussion, regarding the spatial distribution 
of the particles of each species, based on the correlation functions [Fig. \ref{abb:coh3-1}]. 
For instance, $\mathcal{A}^B(\tilde{x'}|\mathcal{A}^A(\tilde{x});t_2)$ shows a population of a right well filament [Fig. \ref{abb:shots3-1} (e)] which does not 
contradict the analysis obtained from the correlation function that a possible particle configuration is the $(A-A-A-B)$. 
Furthermore, we should emphasize that a direct correspondence between the one-body density and one single-shot image is not possible 
due to the small particle number of the considered FF mixture, $N_A=3$ and $N_B=1$, and the presence of multiple orbitals in the system. 
In particular, the MB state is a superposition of multiple orbitals [see Eqs. (\ref{Eq:WF}) and (\ref{Eq:SPFs})] 
and thus imaging an atom alters the MB state of the remaining atoms and consequently the relevant one-body density. 
For a more elaborated discussion on this topic see \cite{mistakidis2017correlation,katsimiga2017many,katsimiga2018many}. 
To obtain the one-body density of the system we average over several single-shot images for each of the species, 
namely $\bar{\mathcal{A}}^A(\tilde{x};t)=1/N_{shots}\sum_{k=1}^{N_{shots}} 
\mathcal{A}_k^A(\tilde{x};t)$ and $\bar{\mathcal{A}}^B(\tilde{x}^{'}|\mathcal{A}^A(\tilde{x});t)=1/N_{shots}\sum_{k=1}^{N_{shots}} 
\mathcal{A}_k^B(\tilde{x}^{'}|\mathcal{A}^A(\tilde{x});t)$, see also Eq. (\ref{averaging}) in Appendix \ref{single_shots_details}. 
In particular, Figs. \ref{abb:shots3-1} (b)-(c) and (f)-(g) present $\bar{\mathcal{A}}^A(\tilde{x};t)$ and 
$\bar{\mathcal{A}}^B(\tilde{x}^{'}|\mathcal{A}^A(\tilde{x});t)$ at time instants $t=t_1$ and $t=t_2$ for an increasing 
number of single-shots $N_{shots}$. 
It becomes evident that upon increasing $N_{shots}$ the averaged images, $\bar{\mathcal{A}}^A(\tilde{x};t)$ 
and $\bar{\mathcal{A}}^B(\tilde{x}^{'}|\mathcal{A}^A(\tilde{x});t)$, display progressively the actual profile of the 
one-body density $\rho^{(1)}_{A}(x)$ and $\rho^{(1)}_{B}(x)$ obtained within ML-MCTDHX [Figs. \ref{abb:shots3-1} (d) and (h)].

\section{Interaction Quench Dynamics of a Particle Balanced Mixture}\label{sec:particle_balanced}

\begin{figure}
	\includegraphics[trim=0 0 0 0,clip,width=0.48\textwidth]{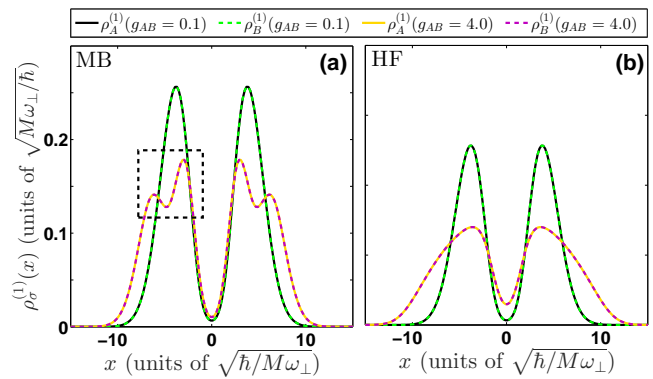}
	\caption{One-body densities $\rho^{(1)}_\sigma(x)$ of the $\sigma$-species ground state of a FF mixture for different 
	interspecies interaction strengths $g_{AB}$ (see legend) within (a) the MB approach and (b) the HF approximation. 
	The mixture consists of $N_A=N_B=2$ fermions and is trapped in a double-well potential. 
	The rectangle in (a) indicates the intrawell fragmentation (filamentation) of $\rho^{(1)}_\sigma(x)$ for strong interactions.}
	\label{abb:gs2-2} 
\end{figure}

\subsection{Ground state}\label{in_par_bal} 

To further elaborate on the interaction quench dynamics of FF mixtures trapped in a double-well potential we next examine particle balanced systems. 
In particular, we study a FF mixture with $N_A=N_B=2$ fermions and follow the same quench scenario as in the above Section \ref{sec:particle_imbalanced}. 
To this end, we first obtain the ground state of the system described by the Hamiltonian of Eq. (\ref{Hamilt}) with interspecies interaction $g_{AB}=0.1$. 
The dynamics is subsequently induced by performing an interaction quench to the strongly interacting regime $g_{AB}=4.0$. 
The double-well possesses a frequency $\omega=0.1$, barrier height $V_0=2$ and width $w=1$. 

The corresponding single-particle density $\rho_\sigma^{(1)}(x;t)$ of the $\sigma$-species ground state [see also Eq. (\ref{one_body_cor})] of the FF mixture is shown in 
Figs. \ref{abb:gs2-2} (a) and (b) for different interspecies repulsions, $g_{AB}$, for both the MB and the HF approach. 
Note that since the FF mixture is particle balanced both equal mass species exhibit exactly the same one-body density, i.e. $\rho^{(1)}_A(x;t)=\rho^{(1)}_B(x;t)$, 
for both approaches. 
For weak interactions, $\rho_\sigma^{(1)}(x;t)$ populate the two wells in a symmetric manner (with respect to reflections at $x=0$), while $\rho_A^{(1)}(x;t)$ 
and $\rho_B^{(1)}(x;t)$ are miscible in both approaches, see Figs. \ref{abb:gs2-2} (a) and (b). 
Turning to the strong interspecies interaction regime we observe that a broadening of $\rho_\sigma^{(1)}(x;t)$ occurs in the HF approximation as a result of the enhanced repulsion. 
In contrast within the MB approach $\rho_\sigma^{(1)}(x;t)$ besides being broadened shows an intrawell fragmentation, indicated by the dashed rectangle in Fig. \ref{abb:gs2-2} (a). 
Let us also mention at this point that a symmetry breaking of $\rho_\sigma^{(1)}(x;t)$ by means of the Stoner instability does not take place in the current setup, 
since both species, besides being mass balanced, contain the same number of particles $N_A=N_B$ and therefore exhibit exactly the same behavior.  
As a consequence phase separation between the species is not favored on the one-body level. 
To induce the nonequilibrium dynamics, in the following, we quench the interspecies repulsion from $g_{AB}=0.1$ to $g_{AB}=4.0$. 

\begin{figure}
	\includegraphics[trim=0 0 0 0,clip,width=0.49\textwidth]{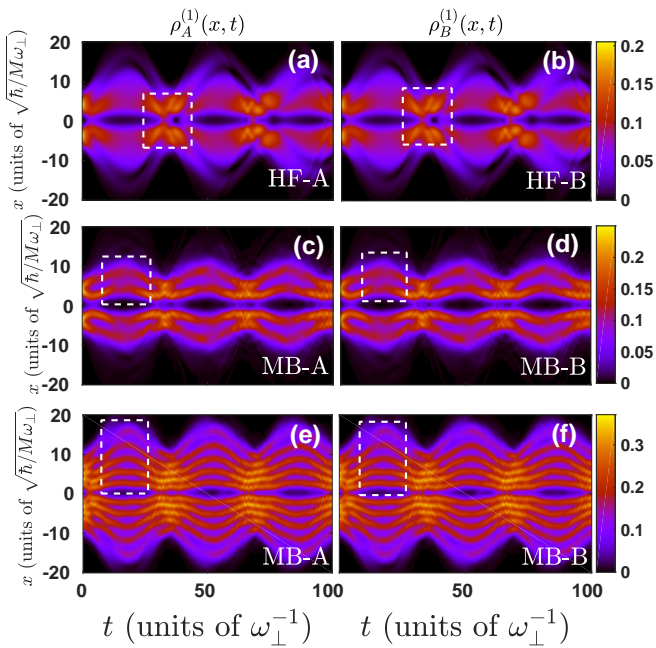}
	\caption{Time-evolution of the one-body density $\rho^{(1)}_\sigma(x;t)$ for the $\sigma$-species for the $A$- (left column) and the $B$-species (right column) 
	of the FF mixture within the (a), (b) HF approximation and (c)-(f) MB approach following an interaction quench from $g_{AB}=0.1$ to $g_{AB}=4.0$. 
	The FF mixture consists of (a)-(d) $N_A=N_B=2$ atoms and (e), (f) $N_A=N_B=5$ fermions. 
	At $t=0$ it is prepared in the ground state of the double-well for $g_{AB}=0.1$. 
	The rectangles in (a), (b) indicate a contraction event of the fermionic cloud and the resulting interference patterns, while in (c)-(f) they mark the 
	number of filaments formed in the left well during the evolution. }
	\label{abb:den2-2} 
\end{figure}

\subsection{Evolution on the Single-Particle level}\label{den_par_bal}

The spatially resolved quench dynamics of the FF mixture can be investigated via the $\sigma$-species one-body density $\rho^{(1)}_\sigma(x;t)$, 
see Figs. \ref{abb:den2-2} (a)-(d). 
Within the HF approximation the quenched one-body density evolution [Figs. \ref{abb:den2-2} (a) and (b)] exhibits 
an overall breathing motion comprising both wells. 
This breathing motion of the fermionic cloud is, of course, characterized by an expansion and contraction of the symmetric 
density branches located in each well. 
Notice that during the contraction process these density branches collide on top of the barrier, i.e. at $x=0$, giving rise to several 
interference patterns [see the dashed rectangles in Figs. \ref{abb:den2-2} (a) and (b)]. 
These interference patterns become even more pronounced for stronger interactions (not shown here). 
Inspecting $\rho^{(1)}_\sigma(x;t)$ within the MB approach, Figs. \ref{abb:den2-2} (c) and (d), we observe that both species undergo a 
breathing mode comprising the double-well but most importantly an intrawell fragmentation of the fermionic cloud takes place within each well. 
In particular, for the $N_A=N_B=2$ case two filaments appear in each well, see here the dashed rectangles in Figs. \ref{abb:den2-2} (c), (d). 
It is worth mentioning at this point that the existence of these filaments is a consequence of beyond HF correlations 
that built in the system \footnote{For instance, inspecting the corresponding orbital densities we observe that the filaments building 
upon higher than the second populated orbitals used (not shown here) are much more pronounced than those developed in the first two orbitals.} 
To conclude upon the dependence of the above-described MB dynamics on the number of fermions in particle balanced FF mixtures, 
Figs. \ref{abb:den2-2} (e) and (f) present $\rho^{(1)}_\sigma(x;t)$ for the same quench amplitude as before (i.e. from $g_{AB}=0.1$ to $g_{AB}=4.0$) 
but for a system containing $N_A=N_B=5$ fermions. 
$\rho^{(1)}_\sigma(x;t)$ possesses a broader distribution when compared to the $N_A=N_B=2$ case and performs an overall breathing motion 
with the same frequency as in the case $N_A=N_B=2$. 
Strikingly enough, the emergent intrawell fragmentation results in five distinct filaments [see the rectangles in Figs. \ref{abb:den2-2} (e), (f)] of 
the $\sigma$-species fermionic cloud within each well. 
Therefore, we can infer that the number of filaments formed $N_f$ is proportional to the particle number $N_f=N_\sigma$. 
We note that we have checked this conclusion also for other particle numbers, e.g. $N_{\sigma}=3,6$ (results not shown here). 

\subsection{Correlation Properties}\label{cor_par_bal}

\begin{figure}
	\includegraphics[trim=0 140 30 0,clip,width=0.5\textwidth]{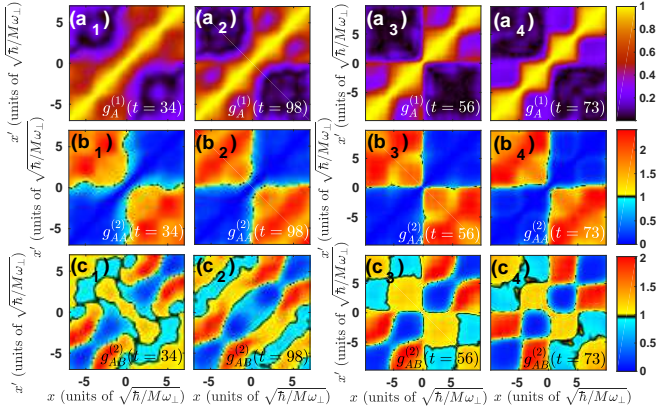}
	\caption{($a_1$)-($a_4$) One-body correlation function $g_A^{(1)}(x,x^\prime;t)$ of the $A$-species for different time instants (see legends) during the 
	interaction quench dynamics from $g_{AB}=0.1$ to $g_{AB}=4.0$ of the particle balanced FF mixture. 
	($b_1$)-($b_4$) The corresponding intraspecies two-body correlation function $g_{AA}^{(2)}(x,x^\prime;t)$ and ($c_1$)-($c_4$) the interspecies 
	two-body correlation function $g_{AB}^{(2)}(x,x^\prime;t)$.  
	The mixture consists of $N_A=N_B=2$ fermions and is initially prepared in the weakly interacting, $g_{AB}=0.1$, ground state of the double-well.}
	\label{abb:coh2-2} 
\end{figure}

In order to expose the role of correlations in the above-discussed interaction quench dynamics of particle balanced FF mixtures we resort to 
the corresponding $g_{\sigma}^{(1)}(x,x^\prime;t)$ [Eq. (\ref{one_body_cor})] and the $g_{\sigma\sigma'}^{(2)}(x,x';t)$ [Eq. (\ref{two_body_cor})] 
correlation functions, see Fig. \ref{abb:coh2-2}. 
Since the considered FF mixture is particle balanced it holds that $g_A^{(1)}(x,x^\prime;t)= g_B^{(1)}(x,x^\prime;t)$ 
and $g_{AA}^{(2)}(x,x^\prime;t)=g_{BB}^{(2)}(x,x^\prime;t)$. 
To this end, below we discuss only $g_A^{(1)}(x,x^\prime;t)$, $g_{AA}^{(2)}(x,x^\prime);t$ and $g_{AB}^{(2)}(x,x^\prime;t)$ following an 
interaction quench of the $N_A=N_B=2$ FF mixture from $g_{AB}=0.1$ to $g_{AB}=4.0$.  

Inspecting $g_A^{(1)}(x,x^\prime;t)$ we deduce that each filament is fully coherent with itself throughout the evolution, since    
$g_A^{(1)}(x,x^\prime\approx x;t)\approx1$ (with $x$ varying on the spatial scale of each filament) as shown in Figs. \ref{abb:coh2-2} (a\textsubscript{1}-(a\textsubscript{4}). 
Regarding the coherence of two distinct filaments we discern between the cases of expansion [e.g. at $t_1=34$, $t_2=98$ in 
Figs. \ref{abb:den2-2} (c), (d)] and contraction [e.g. at $t_3=56$, $t_4=73$ in Figs. \ref{abb:den2-2} (c), (d)] of the fermionic cloud. 
Referring to contraction events, see Figs. \ref{abb:coh2-2} (a\textsubscript{1}) and (a\textsubscript{2}), we observe that two filaments 
residing in distinct wells are fully incoherent between each other [see e.g. $g_A^{(1)}(x=-5.5,x^\prime=2.5;t=34)\approx0$ in 
Fig. \ref{abb:coh2-2} (a\textsubscript{1})]. 
However, two filaments located within the same well are partially coherent, see for instance $g_A^{(1)}(x=-5.5,x^\prime=-2.5;t=34)\approx0.5$ 
in Fig. \ref{abb:coh2-2} (a\textsubscript{1}). 
In contrast, during the expansion of the cloud, see Figs. \ref{abb:coh2-2} (a\textsubscript{3}) - (a\textsubscript{4}), every two filaments 
independently of their location appear to be fully incoherent among each other [e.g. $g_A^{(1)}(x=-7.0,x^\prime=-2.0;t=56)\approx0.1$ 
in Fig. \ref{abb:coh2-2} (a\textsubscript{3}))]. 
Concluding we can infer that during the particle balanced FF quench dynamics Mott-like one-body correlations \cite{sherson2010single, larson2008mott,katsimiga2017dark} emerge 
either for filaments located at distinct wells (contraction events of the femionic cloud) or for all 
filaments formed (expansion events of the fermionic cloud). 

To gain insight into the two-body character of the dynamics we first study the second order intraspecies correlation function $g_{AA}^{(2)}(x,x^\prime);t$ depicted 
in Figs. \ref{abb:coh2-2} (b\textsubscript{1})-(b\textsubscript{4}). 
Overall, strong anti-correlations occur within each well [e.g. $g_{AA}^{(2)}(x=-5.5,x^\prime=-5.5;t=34)\approx0$ in Fig. \ref{abb:coh2-2} (b\textsubscript{1})], whilst 
between the different wells a correlated behavior takes place [e.g. $g_{AA}^{(2)}(x=-5.5,x^\prime=2.5;t=34)\approx1.3$ in Fig. \ref{abb:coh2-2} (b\textsubscript{1})] in the 
course of the evolution. 
Moreover, the probability of two fermions of the same species to populate non-symmetric (with respect to $x=0$) filaments is favored 
when compared to the probability of occupying symmetric ones [compare $g_{AA}^{(2)}(x=-7.5,x^\prime=2.5;t=56)\approx2.4$ and 
$g_{AA}^{(2)}(x=-2.5,x^\prime=2.5;t=56)\approx1.8$ respectively in Fig. \ref{abb:coh2-2} (b\textsubscript{3})].   

To obtain a further understanding of the FF mixture dynamics, we finally inspect the interspecies correlation 
function $g_{AB}^{(2)}(x,x^\prime;t)$, see Figs. \ref{abb:coh2-2} (c\textsubscript{1})-(c\textsubscript{4}). 
A correlation hole emerges on the diagonal elements of $g_{AB}^{(2)}(x,x^\prime;t)$ [see e.g. $g_{AB}^{(2)}(x=-2.5,x^\prime=-2.5;t=56)\approx0.1$ 
in Fig. \ref{abb:coh2-2} (c\textsubscript{3})] indicating that fermions of different species can not populate the same filament. 
Furthermore, two filaments within the same well are found to be strongly correlated [e.g. $g_{AB}^{(2)}(x=-7.5,x^\prime=-2.5;t=56)\approx2.0$ 
in Fig. \ref{abb:coh2-2}(c\textsubscript{3})] which means that it is likely to be occupied by fermions of $A$ and $B$-species. 
On the other hand, regarding filaments located at different wells it is more preferable for two fermions of different species to reside 
in symmetric (with respect to $x=0$) filaments [$g_{AB}^{(2)}(x=-2.5,x^\prime=2.5;t=56)\approx1.2$ in Fig. \ref{abb:coh2-2} (c\textsubscript{1})] rather than 
non-symmetric ones [$g_{AB}^{(2)}(x=-7.5,x^\prime=2.5;t=56)\approx0.8$ in Fig. \ref{abb:coh2-2} (c\textsubscript{3})]. 
Combining the knowledge gained from the one- and two-body correlations we can conclude that the spatial distribution of the system in terms of the 
$A$- and $B$-species fermions in the four emerging filaments is either $A-B-A-B$ or $B-A-B-A$. 
The above is a few-body precursor of anti-ferromagnetic order.  

\begin{figure}
	\includegraphics[width=0.49\textwidth]{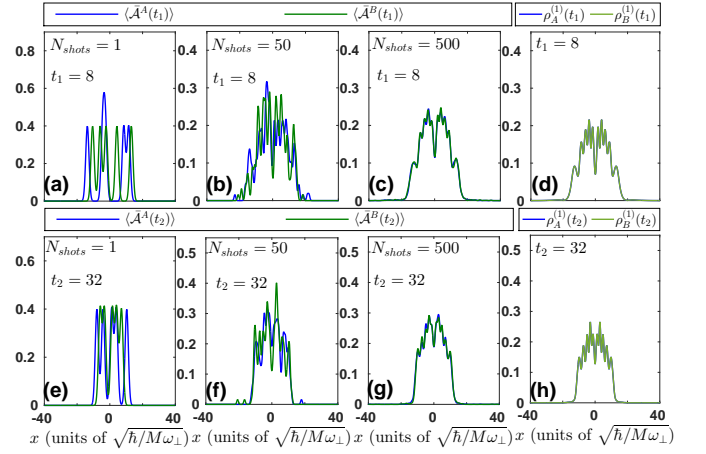}
	\caption{Single-shot images of each species, at distinct time instants (see legends) of the interaction quench dynamics from $g_{AB}=0.1$ to $g_{AB}=4.0$, 
	obtained by averaging over (a), (e) $N_{shots}=1$, (b), (f) $N_{shots}=50$ and (c), (g) $N_{shots}=500$. 
	(d), (h) The corresponding $\rho^{(1)}_{\sigma}(t)$ calculated within the MB approach. 
	In all cases the system consists of $N_A=N_B=5$ fermions trapped in a double-well potential.} 
	\label{abb:shots2-2} 
\end{figure}

\subsection{Single-Shot Simulations}\label{shots_par_bal}

To showcase further experimental links of the above-discussed MB quench dynamics of particle balanced FF mixtures we next briefly discuss 
the outcome of the corresponding single-shot simulations. 
Notice that more details of this procedure can be found in Sec. \ref{shots_par_imb}, in Appendix \ref{single_shots_details} 
and in \cite{mistakidis2017correlation,erdmann2018correlated,koutentakis2018probing}. 
In particular, we consider the setup containing $N_A=N_B=5$ fermions in each component and perform single-shot simulations during the dynamics induced by a 
quench from $g_{AB}=0.1$ to $g_{AB}=4.0$, see also Figs. \ref{abb:den2-2} (e), (f). 
The first single-shot images for both species at $t_1=8$ and $t_2=32$ [see Figs. \ref{abb:shots2-2} (a) and (e)] resemble 
the filamentized structure of $\rho_\sigma^{(1)}(x;t)$ [see Figs. \ref{abb:shots2-2} (a) and (e)] within the double-well 
at both time instants. 
However, as also discussed in Sec. \ref{shots_par_imb}, an adequate correspondence between a single-shot image and the 
corresponding one-body density is not possible due to the small particle number. 
To capture the structures building upon $\rho_\sigma^{(1)}(x)$ we average over several single-shot realizations depicted 
in Figs. \ref{abb:shots2-2} (b)-(c) and (f)-(g) for each time instant. 
As it can be seen, the averaged images $\bar{\mathcal{A}}^A(\tilde{x};t)$ and 
$\bar{\mathcal{A}}^B(\tilde{x}^{'}|\mathcal{A}^A(\tilde{x});t)$ gradually approach $\rho_\sigma^{(1)}(x;t)$ 
[Figs. \ref{abb:shots2-2} (d) and (h)] as $N_{shots}$ is increased.

\section{Conclusions}\label{sec:conclusion} 

We have investigated the nonequilibrium quantum dynamics of a spin-polarized FF mixture confined in a double-well potential upon quenching the 
interspecies repulsion from the weak to the strong interaction regime and for both particle imbalanced and balanced mixtures.  
Comparing the dynamics within the HF approximation and the MB level enables us to infer about the crucial role of correlations 
on both the one- and two-body level in the course of the dynamics. 
In particular, we reveal a variety of interesting phenomena with MB origin such as phase separation processes, alteration of the Stoner 
instability and filamentation of the single-particle density. 

Regarding the ground state of particle imbalanced species a symmetry breaking of the single-particle density occurs 
for strong interspecies interactions within the HF approximation being related to the Stoner's instability that 
renders the two fermionic clouds immiscible. 
Alteration of this instability is observed at the MB level due to the existence of higher-order correlations rendering the two components miscible and 
leading to a prominent intrawell fragmentation of the one-density. 
To induce the dynamics we suddenly change the interspecies interaction from weak-to-strong values. 
It is found that within the HF approximation the $\sigma$-species single-particle density filamentizes, 
i.e. the initial Gaussian-like density profile breaks into several localized density branches called filaments 
while the two species exhibit a dynamical phase separation. 
In sharp contrast, when correlations are included the filamentation of the one-body density becomes more faint and the two species 
show a miscible behavior on the one-body level. 
To provide further insights into the MB character of the dynamics we utilize the one- and two-body correlation functions. 
On the one-body level Mott-like correlations between the filaments are revealed, indicating their tendency for localization. 
Most importantly, both the intra- and interspecies correlation functions show a correlation hole in their diagonal elements 
suggesting that two fermions of the same or different species can not populate the same filament. 
However, the occurence of strong correlations between two distinct filaments indicates that two fermions of the same or different 
species can reside in distinct filaments. 
It is these observations that unveil the phase separated character of the MB dynamics on the two-body level while consisting 
a precursor of anti-ferromagnetic order. 

Turning our attention to particle balanced FF mixtures and their relevant ground state properties, we are able to showcase that while intrawell 
fragmentation occurs at the MB level within the HF approach only a broadening is present. 
In this case the species remain miscible both for weak and strong interspecies interactions independently of the considered approach. 
Performing an interspecies interaction quench from weak-to-strong coupling we observe that in the HF approximation the two 
fermionic clouds remain miscible throughout the evolution. 
Furthermore, they undergo an overall breathing motion over the double-well while in the course of the contraction events of this motion 
prominent interference patterns appear. 
Within the MB approach the two species are miscible and perform an overall breathing mode. 
Most importantly and in sharp contrast to the HF approximation, the clouds exhibit an intrawell fragmentation (filamentation) visible in their  
single-particle density with the number of filaments formed being proportional to the number of fermions of each species. 
Inspecting the one-body correlation function in the course of the evolution we deduce that Mott-like one-body 
correlations appear either for filaments located at distinct wells (contraction events) or for all filaments 
(expansion events of the fermionic cloud). 
Referring to the two-body correlations we find that two fermions of the same or different species exhibit an anti-correlated 
behavior in a single filament while they are strongly correlated when residing in distinct filaments with the non-symmetric ones 
(with respect to the center) being more favorable. 
The above indicate that the two species phase separate suggesting the formation of anti-ferromagnetic like order in the few-body system.  

Finally, we provide possible experimental realizations for both the particle imbalanced as well as the particle balanced cases by 
simulating single-shot measurements. 
In particular, we show how an averaging process of the obtained in-situ images can be used to adequately retrieve the MB fermionic 
quench dynamics.  

There are several promising research directions that are of interest for future investigations 
along the lines of the current effort. 
An imperative prospect is to simulate the corresponding radiofrequency spectrum \cite{mistakidis2018repulsive} in the case 
of particle imbalanced FF mixtures in order to reveal possibly emerging polaronic states and 
subsequently examine their properties. 
Another straightforward direction in particle imbalanced setups would be to consider a larger particle 
number for the minority species, e.g. $N_A=5$, $N_B=3$, and unveil whether phase separation processes and 
magnetization effects occur in such systems. 
Certainly the study of the interspecies interaction quench dynamics of mass imbalanced 
FF mixtures in order to induce a dynamical phase separation of the two species and showcase 
the role of correlations is an intriguing perspective. 


\appendix

\section{The Single-Shot Algorithm} \label{single_shots_details}

The numerical simulation of the single-shot procedure relies on a sampling of 
the MB probability distribution \cite{sakmann2016single,katsimiga2017many,katsimiga2018many,mistakidis2017correlation}. 
We remark that the implementation of this experimental measurement process has already been reported for 
single-component bosons and fermions \cite{sakmann2016single,katsimiga2017many,katsimiga2018many,koutentakis2018probing,erdmann2018correlated} 
as well as for binary bosonic and fermionic mixtures \cite{mistakidis2017correlation}. 
Below we provide a brief sketch of the corresponding numerical procedure but for more details we refer 
the interested reader to \cite{sakmann2016single,katsimiga2017many,katsimiga2018many,koutentakis2018probing,erdmann2018correlated}. 

The single-shots for binary mixtures depend strongly on the system specific inter- and 
intraspecies correlations \cite{mistakidis2017correlation}. 
For a MB system the presence of entanglement [see Eq. (\ref{Eq:WF})] between the species 
is very important for the image ordering. 
In the following, we analyze the corresponding numerical process when the imaging is performed first on the $A$ and then 
to the $B$ species providing this way the absorption images $\mathcal{A}^A(\tilde{x})$ and 
$\mathcal{A}^B(\tilde{x}'|\mathcal{A}^A(\tilde{x}))$. 
An important remark here is that in order to image first the $B$ and then the $A$ species we need to follow the same 
procedure, obtaining the corresponding images $\mathcal{A}^B(\tilde{x})$ and 
$\mathcal{A}^A(\tilde{x}'|\mathcal{A}^B(\tilde{x}))$. 

To achieve the imaging of the $A$ and subsequently of the $B$ species we sequentially annihilate 
all $A$-species fermions. 
In particular at a certain time instant of the imaging, for instance $t_{im}$, a random position is drawn satisfying 
$\rho_{N_A}^{(1)}(x_1')>q_1$ with $q_1$ being a random number belonging to the interval 
[$0$, $ \max\lbrace{\rho^{(1)}_{N_A}(x;t_{im})\rbrace}$]. 
To proceed, the ($N_A+N_B$)-body wavefunction is projected onto the ($N_A-1+N_B$)-body one by the virtue of the 
projection operator $\frac{1}{\mathcal{N}}(\hat{\Psi}_A(x_1')\otimes \hat{\mathbb{I}}_B)$. 
Here $\hat{\Psi}_A(x_1')$ denotes the fermionic field operator annihilating an $A$ species fermion at position $x_1'$, 
while $\mathcal{N}$ is the normalization constant. 
As it can be easily deduced, this process directly affects the Schmidt weights, $\lambda_k$. 
In this way both $\rho^{(1)}_{N_A-1}(t_{im})$ and $\rho^{(1)}_{N_B}(t_{im})$ are changed. 
Recall that the $B$ species have not been imaged yet. 
Indeed, the Schmidt decomposition of the MB wavefunction after this first measurement reads
\begin{equation}
\begin{split}
&\ket{\tilde{\Psi}_{MB}^{N_A-1,N_B}(t_{im})}=\\ &\sum_i \sqrt{\tilde{\lambda}_{i,N_A-1}(t_{im})}\ket{\tilde{\Psi}_{i,N_A-1}^A(t_{im})}\ket{\Psi_i^B(t_{im})}.   
\label{Eq:A1}
\end{split}
\end{equation} 
The $N_A-1$ species wavefunction is $\ket{\tilde{\Psi}_{i,N_A-1}^A}=\frac{1}{N_i}\hat{\Psi}_A(x_1')\ket{\Psi_i^A}$, 
and the normalization factor $N_i=\sqrt{\bra{\Psi_i^A}\hat{\Psi}_A^{\dagger}(x_1')\hat{\Psi}_A(x_1')\ket{\Psi_i^A}}$. 
Also the Schmidt coefficients of the ($N_A-1+N_B$)-body wavefunction read 
$\tilde{\lambda}_{i,N_A-1}=\lambda_i N_i/\sum_i \lambda_i N_i^2$. 
To complete the imaging process we repeat the above steps $N_A-1$ times and then obtain the distribution of 
positions ($x'_1$, $x'_2$,...,$x'_{N_A-1}$). 
The latter is subsequently convoluted with a point spread function resulting in the single-shot 
image of the $A$-species $\mathcal{A}^A(\tilde{x})=\frac{1}{\sqrt{2\pi}w_{PSF}}\sum_{i=1}^{N_A}e^{-\frac{(\tilde{x}-x'_i)^2}{2w_{PSF}^2}}$.  
In this expression $\tilde{x}$ denote the spatial coordinates within the image and $w_{PSF}$ is the width of the employed 
point spread function. 

The MB wavefunction after annihilating all $N_A$ fermions reads 
\begin{equation}
\begin{split}
&\ket{\tilde{\Psi}_{MB}^{0,N_B}(t_{im})}=\\ &\ket{0^A} \otimes\sum_i \frac{\sqrt{\tilde{\lambda}_{i,1}(t_{im})}
\braket{x'_{N_A}|\Phi_{i,1}^A}}{\sum_j{\sqrt{\tilde{\lambda}_{j,1}(t_{im})|\braket{x'_{N_A}|\Phi_{j,1}^A}|^2}}}\ket{\Psi_i^B(t_{im})}.   
\label{Eq:A3}
\end{split}
\end{equation} 
In this expression $\braket{x'_{N_A}|\Phi_{j,1}^A}\equiv\braket{0^A|\hat{\Psi}_A(x'_{N_A})|\Phi_{j,1}^A}$ is the single-particle orbital of the $j$-th mode, while 
the $B$-species wavefunction, i.e. $\ket{\Psi_{MB}^{N_B}(t_{im})}$, is the second term in the cross product 
of the right-hand side. 
As it can be seen, $\ket{\Psi_{MB}^{N_B}(t_{im})}$ refers to a non-entangled $N_B$-particle wavefunction. 
Therefore the subsequent single-shot procedure of the $B$ species is the same as for a single-species 
ensemble \cite{sakmann2016single,katsimiga2017many,katsimiga2018many}. 
This proccedure has been extensively tested for different single-component setups, 
see for more details \cite{sakmann2016single,katsimiga2017many,katsimiga2018many} and references therein. 
Therefore we only brief discuss it below. 
Referring to $t=t_{im}$ i.e. the imaging time, we compute $\rho^{(1)}_{N_B}(x;t_{im})$ 
from $\ket{\Psi^{N_B}_{MB}}\equiv \ket{\Psi(t_{im})}$ and a random position $x''_1$ is drawn obeying $\rho^{(1)}_{N_B}(x''_1;t_{im})>q_2$, 
where $q_2$ is a random number in the interval [$0$, $\rho^{(1)}_{N_B}(x;t_{im})$]. 
Consequently, one particle is annihilated at $x''_1$ and we calculate $\rho^{(1)}_{N_B-1}(x;t_{im})$ 
from $\ket{\Psi^{N_B-1}_{MB}}$. 
Then, a new random position $x''_2$ is drawn from $\rho^{(1)}_{N_B-1}(x;t_{im})$. 
Repeating the above procedure $N_B-1$ times we obtain the distribution of positions 
($x''_1$, $x''_2$,...,$x''_{N_B-1}$). 
This distribution is finally convoluted with a point spread function providing a 
single-shot image $\mathcal{A}^B(\tilde{x'}|\mathcal{A}^A(\tilde{x}))$. 

Finally, it can be shown that the average image of the $\sigma$ species, i.e. $\mathcal{\bar{A}}^{\sigma}(\tilde{x})$, 
over several ($N_{shots}$) single-shot images [$\mathcal{A}^{\sigma}(\tilde{x})$] is directly related to the $\sigma$ species 
one-body density, $\rho_{\sigma}^{(1)} (x_{\sigma}')$, since  
\begin{equation}
 \mathcal{\bar{A}}^{\sigma}(\tilde{x})=\frac{N_{\sigma}}{\sqrt{2\pi}w_{PSF}}\int dx_{\sigma}' e^{-\frac{(\tilde{x}-x_{\sigma}')^2}{2w^2_{PSF}}} \rho_{\sigma}^{(1)} (x_{\sigma}').
 \label{averaging}
\end{equation}
Here, $\tilde{x}$ denote the spatial coordinates within the image and $x_{\sigma}'$ is the spatial coordinate of the $\sigma$ species.  
Also $w_{PSF}$ is the width of the employed point spread function and $N_{\sigma}$ refers to the particle number of the $\sigma$ species.

\section{Convergence and Further Details of the Many-Body Simulations} \label{sec:numerics}

Let us briefly discuss the ingerdients of our MB simulations and showcase their numerical 
convergence. 
As it has been already argued in Sec. \ref{sec:wfn}, ML-MCTDHX \cite{ML-MCTDHX} is a variational method for solving the time-dependent MB Schr{\"o}dinger equation for 
atomic mixtures consisting either of bosonic \cite{mistakidis2017correlation,katsimiga2017dark,mistakidis2018effective} or 
fermionic \cite{cao2017collective,koutentakis2018probing,erdmann2018correlated,mistakidis2018repulsive} species. 
Within this approach, the MB wavefunction is expanded in terms of a time-dependent variationally optimized MB basis. 
Such a treatment, allows us to take into account the relevant intra- and interspecies correlation 
effects utilizing a computationally feasible basis size. 
In this way, the number of basis states can be significantly reduced as compared to methods which rely on a time-independent basis. 
The latter is achieved by choosing the relevant subspace of the Hilbert space at each time instant of the evolution in a more efficient manner.  

\begin{figure}
	\includegraphics[width=0.4\textwidth]{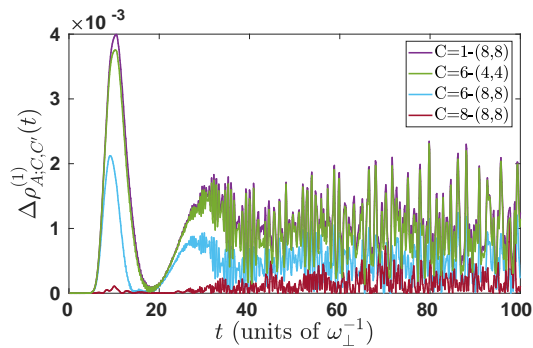}
	\caption{Evolution of the one-body density deviation $\Delta\rho^{(1)}_{A;C,C^\prime}(t)$ between the $C^\prime=10-(8,8)$ and other 
	orbital configurations $C=M-(m^A,m^B)$ (see legend). 
	The FF mixture consists of $N_A=N_B=2$ fermions and to induce the dynamics we perform a quench from $g_{AB}=0.1$ 
	to $g_{AB}=4.0$.}
	\label{abb:convergence} 
\end{figure}

The Hilbert space truncation refers to the employed numerical configuration space designated by $C=D-(m^A,m^B)$. 
In this notation $D=D^A=D^B$ and $m^A$, $m^B$ correspond to the number of species and single-particle functions respectively 
for each of the species [see also Eqs. (\ref{Eq:WF}) and (\ref{Eq:SPFs})]. 
For our simulations we invoke a primitive basis based on a sine discrete variable representation 
including 400 grid points. 
To conclude upon the convergence of our MB simulations we assure that variations of 
the numerical configuration space $C=D-(m^A,m^B)$ do not essentially affect the observables of interest. 
Note that all MB calculations presented in the main text are based on the numerical configuration space $C=6-(6,6)$ for $N_A=3$, $N_B=1$, 
on the $C=10-(10;10)$ in the case of $N_A=5$, $N_B=1$ and $N_A=N_B=5$ and on the $C=10-(8;8)$ when $N_A=N_B=2$. 
Therefore, the available Hilbert space for the corresponding simulation includes 4992 (10720) and 13140 (7060) coefficients for the 
$N_A=3$, $N_B=1$ ($N_A=5$, $N_B=1$) and the $N_A=N_B=5$ ($N_A=N_B=2$) cases respectively. 
This is in sharp contrast to an exact diagonalization procedure which should take into account 4.2 $10^9$ (3.3 $10^{13}$) and 6.9 $10^{21}$ (6.3 $10^{9}$) 
coefficients for the $N_A=3$, $N_B=1$ ($N_A=5$, $N_B=1$) and the $N_A=N_B=5$ ($N_A=N_B=2$) cases, rendering 
these simulations infeasible.  

Finally, let us briefly showcase the convergence of our results for a varying number of species and single-particle functions. 
For this investigation we resort to the $\sigma$-species one-body density, $\rho^{(1)}_{\sigma;C}(x;t)$, during the nonequilibrium dynamics 
and calculate its spatially integrated absolute deviation for each of the species between the $C'=10-(8,8)$ and other numerical configurations 
$C=D-(m^A,m^B)$. 
Namely 
\begin{equation}
\Delta\rho^{(1)}_{\sigma;C,C^\prime}(t) =\frac{1}{2N_\sigma}\int dx|\rho_C^{(1),\sigma}(x;t) -\rho_{C^\prime}^{(1),\sigma}(x;t)| \label{convergence} 
\end{equation} 

Figure \ref{abb:convergence} presents $\Delta\rho^{(1)}_{A;C,C^\prime}(t)$ for a FF mixture with $N_A=N_B=2$ fermions following an interspecies interaction quench 
from $g_{AB}=0.1$ to $g_{AB}=4.0$. 
We remark, that $\Delta\rho^{(1)}_{B;C,C^\prime}(t)=\Delta\rho^{(1)}_{A;C,C^\prime}(t)$ at all times and for all configurations due to the particle 
balanced mixture. 
Therefore the results of the $A$-species are representative for both species. 
Inspecting Fig. \ref{abb:convergence}, it becomes evident that a systematic convergence of $\Delta\rho^{(1)}_{\sigma;C,C^\prime}(t)$ can be achieved. 
More specifically, comparing $\Delta\rho^{(1)}_{A;C,C^\prime}(t)$ between the $C=8-(8,8)$ and $C'=10-(8,8)$ approximations we observe that the corresponding 
relative difference is below $0.15\%$ throughout the evolution. 
Finally, we remark that a similar analysis has been performed for all other particle configurations, i.e. particle number imbalanced systems as well 
as higher particle numbers, discussed within the main text and found to be adequately converged (not shown here for brevity).

\section*{Acknowledgements} 

S.I.M. and P.S. gratefully acknowledge financial support by the Deutsche Forschungsgemeinschaft (DFG) in the framework of the 
SFB 925 ``Light induced dynamics and control of correlated quantum systems''. 

\bibliographystyle{apsrev4-1.bst}
\bibliography{bibliography}
\end{document}